\documentstyle[tighten,pra,aps,multicol,epsfig]{revtex}
\begin{document}

\title{Mixed State Entanglement: \\ Manipulating
Polarisation-Entangled Photons}
\author{R.~T.~Thew\cite{RTT} and W. J. Munro\cite{RTT,WJM}}
\address{\cite{RTT}Centre for Quantum Computer Technology, University of
Queensland,\\ QLD 4072, Brisbane, Australia \\ \cite{WJM} Hewlett
Packard Laboratories, Filton Rd, Stoke Gifford, Bristol, BS34 8QZ, UK}
\date{\today}
\maketitle

\begin{abstract}
There has been much discussion recently regarding entanglement
transformations in terms of local filtering operations and whether the optimal
entanglement for an arbitrary two-qubit state could be realised. We introduce an
experimentally realisable scheme for manipulating the entanglement of
an arbitrary state of two polarisation entangled qubits. This scheme
is then used to provide some perspective to the mathematical concepts
inherent in this
field with respect to a laboratory environment. Specifically, we look at how to extract
enhanced entanglement from systems with a fixed rank and in the case
where the rank of the density operator for the state can be reduced,
 show how the state can be made arbitrarily close to a maximally entangled pure
state. In this context we also discuss bounds on entanglement in mixed states.
\end{abstract}

\pacs{03.67.-a, 42.50.-p,03.65.Bz}
\begin{multicols}{2}

\section{Introduction}
   Since the foundations of quantum mechanics were laid, one of its
most curious, and perhaps defining, features has been
entanglement. Historically this has been discussed with regard to questions
of nonlocal
behaviour of quantum systems, a consequence of the famous EPR paper
\cite{EPR} and subsequent work by Bell \cite{Bell}. In the past decade
the focus has shifted to a more information-theoretic interpretation
of entanglement in line with the global effort to understand and
eventually build a quantum computer. Quantum computing is not the only
avenue that has motivated interest. In more immediate terms, realistic
endeavours involve quantum cryptographic schemes, dense coding and
teleportation as well as  general questions regarding quantum information
\cite{Preskill:98,Nielsen:99}. While the realisation of a quantum
computer is a long term goal, these pursuits are motivating an
enormous amount of cross-disciplinary collaboration in questioning
some of the fundamentals of quantum mechanics, information theory, and
how the two are related.

The centerpiece of much of this work is entanglement. Quantifying,
 generating, distributing and maintaining entanglement make up the cornerstones
 of an enormous amount of research in quantum information
 science. A means of manipulating entanglement will
 be vital in distributing and maintaining entanglement, and photons
 provide the most realistic and accessible means of achieving
 this. In this paper we refine an experimentally realisable \cite{Kwiat:00}
 protocol for manipulating arbitrary states of polarisation entangled
 photons which we have previously introduced\cite{Thew:01}. This scheme has
 been significantly improved and we provide here extensive analysis of
 the protocol in the context of entanglement transformations. This scheme specifically targeted mixed states, as
 experimentally it is unrealistic to consider the system isolated from interactions
with the environment. We would also like to connect some of the mathematical concepts regarding entangled mixed states with a more intuitive and realistic experimental
 proposal. In terms of
 manipulating a state's entanglement and purity there was a proposal of
 Kent {\it et al.} \cite{Kent:99} pertaining to the requirements for
 an optimal entanglement transformation. This is all performed in the context of local
 filtering operations and in this paper we will show how and why this works
 in a system using polarisation entangled photons and allowing for
 imperfect photo-detection.

 As only two qubit states are considered, an
exact expression for the {\it Entanglement of Formation} ($EOF$), introduced by Wooters \cite{Wooters:98}, will be used.

The $EOF$ is
\begin{eqnarray}
 E(C(\rho)) & = & h\left(\left[1 + \sqrt{1 - C(\rho)^2}\right]/2\right)
\end{eqnarray}
where $h$ is the binary entropy function,
\begin{equation}
h(x) = -x ~ \log(x) -(1-x)\log(1-x)
\end{equation}
This is derived in terms of the spin flip operation
\begin{equation}
\tilde{\rho} = (\sigma^A_y\otimes\sigma^B_y)\rho^*(\sigma^A_y\otimes\sigma^B_y)
\end{equation}
where $\sigma_y$ are Pauli operators and the complex conjugation is taken
in the computational basis. From this the Concurrence can be found
\begin{eqnarray}
C(\rho) = \max\{\tilde{\lambda}_1 -\tilde{\lambda}_2 -
\tilde{\lambda}_3 - \tilde{\lambda}_4, 0\}
\end{eqnarray}
where the square root of the eigenvalues for
$\rho\tilde{\rho}$, $\tilde{\lambda_i}$, are sorted into descending
order.

 The other characteristic that is considered here is the purity of the
state and the {\it von Neumann Entropy} provides a convenient and
useful measure.  The entropy of the bipartite density matrix, $\rho_{AB}$, is
\begin{eqnarray}
S(\rho_{AB})~ =~ -{\rm Tr}\left[\rho_{AB} \log_4 \rho_{AB}\right]
&  = & -\sum_{i = 1}^{4} \lambda_i \log_4 \lambda_i
\end{eqnarray}
where $\lambda_i$ are the eigenvalues of $\rho_{AB}$. In the latter form
this is analogous to the classical Shannon entropy. The $\log$ to base 4 is used
as this is the joint state, and hence in this
form returns a
normalised entropy ranging from zero, for a pure state, to one for the
Identity or  totally mixed state.

 For a correlated
system the entropy of the whole system is less than the entropy  of its parts due
to the information that is present in the correlations between the two
systems.  For a maximally entangled pure state $S(\rho_A) = S(\rho_B)
= \log(2)$ and
$S(\rho_{AB}) = 0$.  How the state was prepared cannot be determined by
considering measurements on the two subsystems. The correlation in the
joint state measurements must be considered.

 The characteristics of the entropy for a mixed state, regarding both the
joint state
and local subsystem, will be useful when discussing state
transformations. The entropy  provides a key element in discussing bounds
on the Hilbert space associated with mixed states in the context of state
manipulation in general and the scheme introduced here. These concepts
will be discussed primarily in terms of a proposed bound on mixed state entanglement enhancement
\cite{Kent:99} that requires the subsystem entropies to be maximised. 
\section{Entanglement Manipulation using Beam Splitters}
The entanglement manipulation protocol that is to be introduced here
relies on the very simple process of filtering, a method proposed by
several people \cite{Popescu:95,Gisin:96,Horodecki:97a} as a
means of manipulating entangled states. This protocol is conceptually
similar to the {\it Procrustean Method} introduced by Bennett {\it et al.}
\cite{Bennett:96a} which dealt with pure states of entangled spin-half
particles, in a very generic way.

Our aim is for two parties, $A$ and $B$, who are spatially separated,
to share the optimal entanglement available. The qubits we consider
here are polarisation states, where $|V\rangle, |H\rangle$ correspond
to the $|0\rangle, |1\rangle$ states within the standard computational basis.

\begin{figure}[!h]
\epsfig{figure=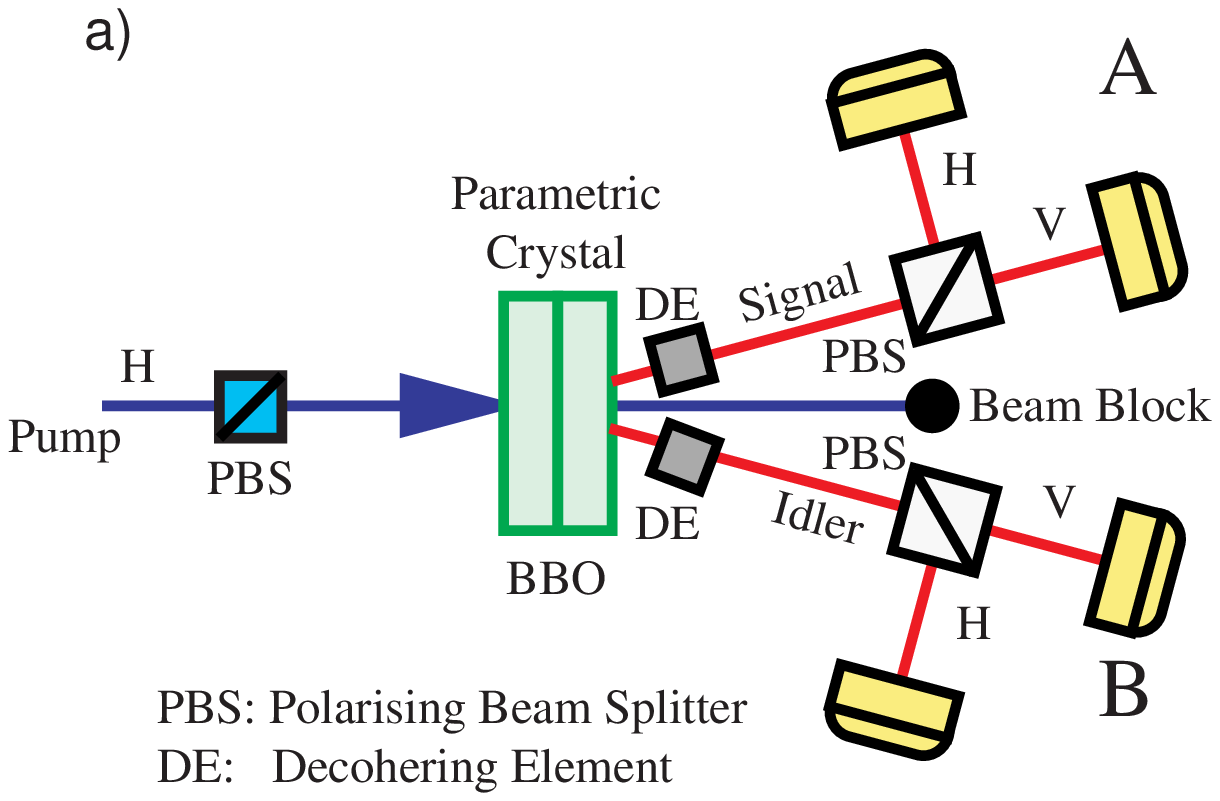,width=85mm} \\
\epsfig{figure=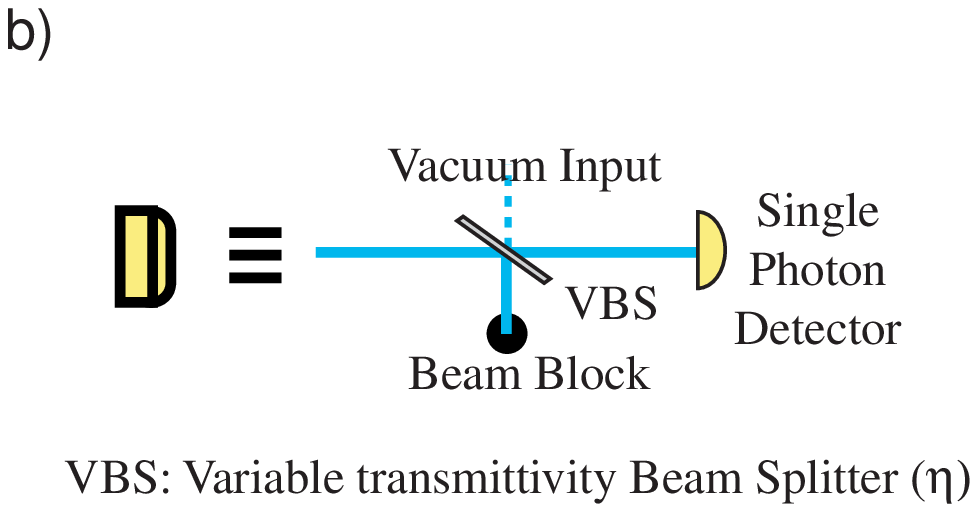,width=70mm}
\caption{\label{fig:pbss} The initial polarised beam is
incident on a Polarising Beam Splitter (PBS) creating general
superposition states which are dependent on the orientation of the
PBS. This beam then undergoes a Down-conversion process at the BBO
crystal producing pure states where the degree of entanglement is
determined from the initial superposition. The Signal and Idler
outputs are then subject to independent Decohering Environments (DE)
allowing variation in the mixedness of the state. This allows the
generation of a wide variety of entangled states. The schematic for
the Beam Splitter Protocol illustrates how an entangled state shared
between $A$ and $B$ is spatially separated with respect to its
polarisation modes by PBSs. Each mode is then incident on another beam
splitter with variable transmittivity (VBS). With some prior knowledge
of the state the VBSs can be manipulated, concentrating the
characteristics of the output state that has coincidence detections at
$A$ and $B$.}
\end{figure}

The schematic in Figure (\ref{fig:pbss}) represents the
proposed manipulation protocol which will be referred to from here on
as the Beam Splitter Protocol. Everything to the left of, and
including, the BBO (Beta-Barium-Borate) crystal and decohering
elements (DEs) are representative of
the source that can supply the initial entangled states that we propose to  manipulate. The first Polarising Beam Splitter (PBS) at the input, before the
crystal, varies the weighting of a superposition state which is then
Down-converted at the Parametric crystal generating pure entangled pairs. The
decohering elements, after the crystal, vary the mixedness of the
state. The recent advances in the preparation
of non-maximally entangled pure \cite{White:99} and mixed \cite{White:00}
polarisation-entangled states allows for the consideration of a wide
variety of initial states, with high production rates for the entangled
photons\cite{Kwiat:99}.

The scheme operates in the following manner. The output from the
crystal, the two arms labelled  {\it signal} and {\it idler}, are incident on  PBSs, spatially separating the vertical and
horizontal polarisation modes of the two separate beams. These
modes will be labelled, $ |V\rangle_A, |H\rangle_A, |V\rangle_B,
|H\rangle_B$. Both the polarisation modes, $V$ and $H$, in both the arms, $A$
and $B$, will then be incident on Variable Beam Splitters (VBSs). These variable
beam splitters can then be adjusted to obtain the desired
output state dependent on the transmission coefficient, $\eta$,
for each mode. This transmission is polarisation dependent. Due to low detector efficiencies, in this protocol the
reflected modes are ignored and
the final state that is considered is the state that has coincidence
detections at both $A$ and $B$. This will be  justified shortly.

 All four Bell-type states will be considered here.  A mixture
of two of these non-maximally entangled Bell-type states,
\begin{eqnarray}
|\phi^{\pm}\rangle_{AB} = \cos \theta_1|VV\rangle_{AB}\pm \sin \theta_1
|HH\rangle_{AB}\label{eq:nmeps1}\\
|\psi^{\pm}\rangle_{AB} = \cos \theta_2|VH\rangle_{AB}\pm \sin \theta_2
|HV\rangle_{AB}\label{eq:nmeps2}
\end{eqnarray}
will be used to illustrate the extension from pure to mixed state
manipulation. The degree of entanglement in each of these states is
determined by the
value of $\theta$, i.e. a maximally entangled state will have equal
weighting  of the
coefficients, $\theta_{1,2} = \pi/4$.

When we consider a beam splitter interaction we must also consider that
as well as the incident mode, the other port of the beam splitter is
subject to the vacuum and similarly the output will also have two modes. The effect that the beam splitters have on a
polarisation-entangled state is to transform the modes  in the
following way
\begin{eqnarray}
 |V,H \rangle_{AB}|0 \rangle \rightarrow \eta_{V,H}|V,H \rangle_{AB} |0 \rangle +
  \sqrt{(1-\eta_{V,H}^2)}|0\rangle_{AB} |1\rangle
\end{eqnarray}
This can be interpreted as the vertical or horizontal modes being passed by
the beam splitter with a probability $\eta_{V,H}^2$, with a component at the
reflected port that now has a photon in an ancilla mode with a
probability $(1 - \eta_{V,H}^2)$. This approach has a similar interpretation to
those found by modelling imperfect detectors as a perfect detector plus a beam
splitter attenuating the input field \cite{Milburn:94}. 

It is easy to determine how a single beam splitter in one arm of the
system could be coupled to a specific vertical or horizontal mode. It
is not much harder to do this for a beam splitter at $A$ and $B$,
however, we wish to introduce two beam splitters, both  vertical and
horizontal, to each polarisation arm of the system. This couples a controllable variable, the
transmission coefficient,  to each mode, where the four variable beam
splitters act independently on the four polarisation modes of the
bipartite system. 

Consider a non-maximally entangled pure state of the form of
eq.(\ref{eq:nmeps2}). After interactions with all four of
the variable beam splitters, the final state, before anything is
discarded, is
\begin{small}
\begin{eqnarray}\label{eq:psm}
|\psi \rangle_{tot} &=&\tilde{{\cal N}} \left\{\right.\left[\cos \theta\eta_{VA}\eta_{HB}|VH\rangle_{AB}\rangle \pm
\sin \theta\eta_{HA}\eta_{VB}|HV\rangle_{AB}  \right] |00\rangle\nonumber \\
&\;&\;\;\;+ \cos \theta \eta_{VA} \sqrt{(1-\eta_{HB}^2)}
\left(|V0\rangle_{AB}|01\rangle + |0H\rangle_{AB}|10\rangle \right) \nonumber \\
&\;&\;\;\; \pm \sin \theta \eta_{HA} \sqrt{(1-\eta_{VB}^2)} \left
(|V0\rangle_{AB}|01\rangle+ |0H\rangle_{AB}|10\rangle \right)\nonumber \\
&\;&\;\;\; + \cos \theta
\sqrt{(1-\eta_{VA}^2)}~\sqrt{(1-\eta_{HB}^2)}|00\rangle_{AB}|11\rangle
\nonumber \\
&\;&\;\;\; \pm \sin \theta
\sqrt{(1-\eta_{HA}^2)}~\sqrt{(1-\eta_{VB}^2)}|00\rangle_{AB}|11 \rangle
\left.\right\}
\end{eqnarray}
\end{small}
The modes can be interpreted as those labeled $AB$ being the transmitted
modes and the others being the ancilla. Also, for convenience,
information regarding  the
polarisation of the photons in the ancilla modes has been discarded
and a simple record of whether there is, or isn't, a photon in a
reflected port at $A$  or $B$, which is
all that is required, has been kept. 

It has been remarked previously
that due to low detector efficiencies the reflected component is
ignored and the state with coincidence detection at $A$ and $B$ is
considered.  In eq.(\ref{eq:psm}) it can be seen that the only
components of the state having coincidence detections at $A$ and $B$
are the two components on the first line. This can be considered to be
a coincidence basis state.  The coincidence basis state is the state that
would have coincidence detections at both $A$ and $B$, i.e.
detections for any of $\{ |VV\rangle_{AB}, |VH\rangle_{AB}, |HV\rangle_{AB},
|HH\rangle_{AB} \}$. Alternatively, if it was at all possible to
efficiently detect single photons, then perfect single-photon
detectors could replace each of the beam blocks in the signal and
idler arms in Figure (\ref{fig:pbss}b). This would allow the system
to operate a
gate-like device at the output that, with the aid of classical
communication between $A$ and $B$,  was open and letting maximally
entangled pairs through as long as a detection is made at one of the
previously discarded ports. Again with reference  to
eq.(\ref{eq:psm}), if this condition was satisfied then the output
state to which $A$ and $B$ would have access corresponds to what has
been referred to as  the coincidence basis state.  As perfect
photodetection is not a realisable process
with current technologies, the beam blocks remain and the state having joint
coincidences for $A$ and $B$ is considered. This  leaves a reduced output state

\begin{eqnarray}
|\psi\rangle_{f} ={\cal N}\left[\eta_{VA}\eta_{HB}\cos \theta|VH\rangle_{AB}
+\eta_{HA}\eta_{VB}\sin \theta|HV\rangle_{AB}
\right]
\end{eqnarray}
with the normalisation
\begin{equation}
{\cal N} = \left[\eta_{VA}^2\eta_{HB}^2\cos^2\theta +
\eta_{HA}^2\eta_{VB}^2\sin^2\theta\right]^{-1/2}
\end{equation}
This is a post-selective operation, selecting a
subensemble with improved entanglement characteristics and discarding the rest of the
ensemble. If no detection is made then
the state can be jointly discarded by both $A$ and $B$. This
post-selective process has the advantage that poor detector
efficiencies only decrease the coincidence count rate. The requirement
for a maximally entangled state is therefore given by
\begin{equation}\label{eq:psc1}
\cos^2\theta ~\eta^2_{VA}\eta^2_{HB} = \sin^2\theta ~\eta^2_{HA}\eta^2_{VB}
\end{equation}
If $\cos\theta > \sin\theta $ then either $\eta_{VA}$ or $\eta_{HB} $, or
both, can be varied such that $\eta^2_{VA}\eta^2_{HB} = \tan^2\theta$,
thus obtaining a maximally entangled state with probability
\begin{equation}
P = 2\sin^2\theta
\end{equation}
which constitutes an optimal transformation for single-copy pure
states \cite{Lo:97,Vidal:99}. The probability of producing the maximally entangled pure state for
this protocol is
dependent on the beam splitter transmission coefficients and is determined
from the trace of the reduced output state density matrix. This is the
probability of obtaining the desired state after the beam splitter
settings are determined.

This provides an intuitively simple explanation of this process, for pure
states at least, but if mixed states are to be considered then a more
convenient representation can be obtained by using the generalised
measurement formalism. This procedure
constitutes a  generalised measurement in that an ancilla is attached
to the system, unitary transformations are performed in the extended
Hilbert space where measurements are made and then
part of the system is traced  out and discarded \cite{Nielsen:97}. 

As we are only interested in the coincidence basis output state, an
equivalent local filtering operation can be derived that retains the
polarisation coupling characteristics derived for the pure state
case. Therefore an  {\it effective} transmission matrix for the joint
system can be written 
\begin{eqnarray}\label{eq:filter}
(A\otimes B) & = &  \left[ \begin{array}{cccc}
 		\eta_{VA}\eta_{VB}   & 0 & 0& 0\\
		0 & \eta_{VA}\eta_{HB}   & 0 & 0 \\
		 0 &  0 & \eta_{HA}\eta_{VB} & 0  \\
             0 & 0 & 0 &  \eta_{HA}\eta_{HB}  		
\end{array} \right]
\end{eqnarray}
 It can easily be seen that this effective transmission matrix allows for the completely positive mapping of the
input state to the coincidence basis output state. The
total state transformation matrix is operating in an Hilbert space
considerably larger than the original state space, and in this expanded
Hilbert space there is now a greater degree of freedom in which to
manipulate the state. This is, in part, where the original Procrustean
Method obtained its name in that it takes an
initial state, places it in an extended Hilbert space, and then manipulates
and discards anything not needed.

Thus with all the transmission coefficients acting independently on the
$\{|V\rangle_A, |H\rangle_A, |V\rangle_B, |H\rangle_B\}$ modes, the
transmission matrix of eq.(\ref{eq:filter}) represents the Beam Splitter
manipulation process. This process is analogous to many of the
filtering operations that have been proposed \cite{Gisin:96,Horodecki:97a}. Any of the beam splitter transmission coefficients
$\eta_{VA}, \eta_{HA}, \eta_{VB}, \eta_{HB}$, can be manipulated
individually or in unison. {\it The key feature of this proposal is that
each polarisation mode in $A$
and B can be manipulated independently}. The degree of freedom that this
protocol provides means that a wide variety of operations for transforming
a bipartite system can be satisfied.

 The output state, or more specifically, the reduced {\it coincidence basis}
output state, can now be written in the form
\begin{eqnarray}\label{eq:operator}
\hat{\rho}_{out} = \frac{A\otimes B~\hat{\rho}_{in}~ A^{\dagger}\otimes
B^{\dagger}}{{\rm Tr}[ A \otimes B ~\hat{\rho}_{in}~ A^{\dagger}\otimes
B^{\dagger}]}
\end{eqnarray}
This state describes the subensemble that passes the filtering process and
would have coincidence detections at $A$ and $B$. The probability of
this state being realised is given by
\begin{eqnarray}
P = {\rm Tr}[A \otimes B~ \hat{\rho}_{in}~ A^{\dagger} \otimes
B^{\dagger}]
\end{eqnarray}
The only restriction on these operations is that they must satisfy
$A^{\dagger}A \le I$ and  $B^{\dagger} B \le I$, being completely
positive maps \cite{Nielsen:97}.

The case of pure states provided a straightforward example of how this
protocol works. So far though, only two of the Bell-type states have been
considered. To illustrate the
transmission matrix method and cover the other Bell state variants, consider
the pure state
\begin{equation}
|\phi^{\pm}\rangle_{in} = \cos \theta|VV\rangle \pm \sin \theta|HH\rangle
\end{equation}
This state has the explicit density matrix representation
\begin{eqnarray}\label{eq:nmedm1}
\hat{\rho}_{in} & = &  \left[ \begin{array}{cccc}
 	\cos^2\theta & 0 & 0 & \pm \cos\theta \sin\theta \\
  0 & 0  & 0 & 0 \\
 0 &  0 & 0 & 0  \\
  \pm \cos\theta \sin\theta & 0 & 0 & \sin^2\theta	\end{array} \right]
\end{eqnarray}
If we apply the Transmission matrix to this state then the output state, given the matrix notation, is
\begin{eqnarray}
\hat{\rho}_{out} & = & \frac{1}{P}\left[ \begin{array}{cccc}
\eta_{VA}^2\eta_{VB}^2\cos^2\theta  & 0 & 0 & \pm
\bar{\eta} \cos\theta \sin\theta \\
             0 & 0 & 0 & 0\\
	 0 &  0 & 0 & 0  \\
 \pm \bar{\eta}\cos\theta \sin\theta  & 0 & 0 &
\eta_{HA}^2\eta_{HB}^2\sin^2\theta  		
\end{array} \right]
\end{eqnarray}
with $\bar{\eta} = \eta_{VA}\eta_{HA}\eta_{VB}\eta_{HB}$, and  the
probability is given by the trace of the unnormalised,
beam-splitter-transformed, density matrix
\begin{eqnarray}
 P & = & \eta_{VA}^2\eta_{VB}^2\cos^2\theta  + \eta_{HA}^2\eta_{HB}^2\sin^2\theta
\end{eqnarray}
A  maximally entangled state is recovered from the coincidence basis
output state
\begin{equation}
|\phi^{\pm}\rangle_{out} =
\frac{1}{\sqrt{P}}\left[\eta_{VA}\eta_{VB}\cos\theta|VV\rangle
\pm \eta_{HA}\eta_{HB}\sin\theta|HH\rangle \right]
\end{equation}
providing  the requirement for a maximally entangled state
\begin{eqnarray}
\cos\theta~ \eta_{VA}\eta_{VB}& = &\sin\theta~ \eta_{HA}\eta_{HB}
 \nonumber \\\nonumber \\
\frac{\eta_{VA}\eta_{VB}}{\eta_{HA}\eta_{HB}}& = &\tan\theta
\end{eqnarray}
are met. If $\cos\theta > \sin\theta $ then either $\eta_{VA}$ or $\eta_{VB}$,
or both, can be varied producing a maximally entangled state with
probability $P = 2\sin^2\theta $. Conversely, if  $\cos\theta <
\sin\theta $ then varying  $\eta_{HA}$ or $\eta_{HB}$, would yield a
maximally entangled state with
probability $P = 2\cos^2\theta $. It could be argued that this constitutes
nothing more than a simple variation on the Procrustean
Method \cite{Bennett:96a} and requires only filtering at either $A$ or $B$ to
distill maximally entangled pure states. The reason for having four
individually tunable filters is perhaps not clear yet, and though there is
obviously a large degree of freedom in controlling the system, the
necessity will become more apparent as the mixed state case is investigated.

\section{Mixed States Manipulation}
\begin{figure}[!h]
\begin{center}
\epsfig{figure=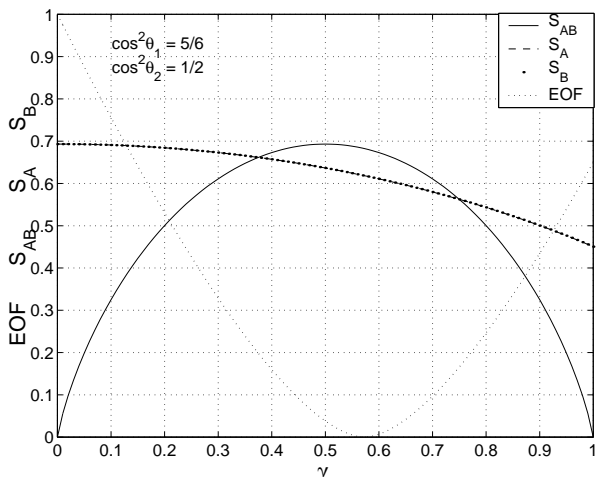,width=80mm}
\caption{\label{fig:tmse} The EOF and Entropy, for the joint state and
local subsystems, as the degree of mixing, $\gamma$, is varied.  The
entanglement decreases as the mixing increases, as expected, between a
maximally entangled state, $\gamma= 0$ and a non-maximally entangled
state, $\gamma= 1$. The subsystem entropies are equal and reach a
maximum of log(2) at $\gamma= 0$.}
\end{center}
\end{figure}

\begin{figure}[h]
\begin{center}
\epsfig{figure=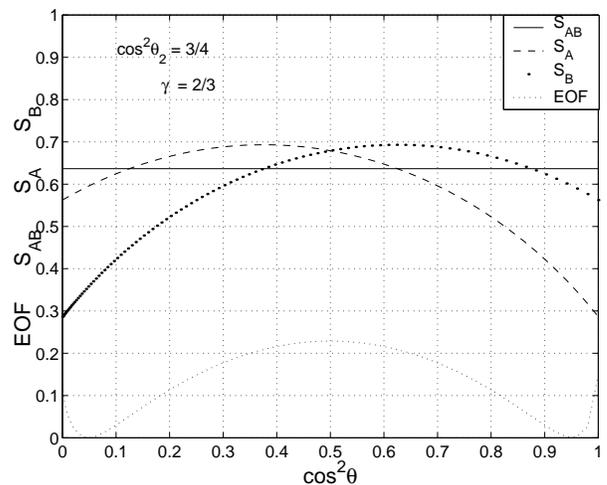,width=80mm}
\caption{\label{fig:tmsea}The EOF, and Entropy, for the
joint state and local subsystems, as the degree of entanglement of one
of the pure state components is varied, for a fixed degree of
mixing. In this instance the subsystem entropies are not always equal,
however where they are corresponds to the point of maximal entanglement
for the joint state.}
\end{center}
\end{figure}

It is the aim of this section to show how the Beam Splitter Protocol can be
extended from pure state manipulation to deal with the more complicated mixed
state manipulation. To aid in the  understanding of how the protocol can
realise this, a state which involves a mixture of two of the non-maximally
entangled pure states already discussed will be be introduced. The degree
of  entanglement of each of the states can be varied as a function of
$\theta_{1,2}$ and the mixing of the two will be determined by another
parameter $\gamma$, such that the state has the density matrix
representation
\begin{eqnarray}\label{eq:tmstate}
\hat{\rho}(\gamma) &= &\gamma |\phi^+\rangle\langle \phi^+| +
(1-\gamma)|\psi^+\rangle\langle\psi^+|
\end{eqnarray}
where $|\phi^+\rangle$ and $|\psi^+\rangle$ correspond to positive
variants of eq.(\ref{eq:nmeps1}) and eq.(\ref{eq:nmeps2}) respectively.
This state will be discussed in terms of the effect that varying the
entanglement and mixing of the components has on the entropy and
entanglement of the system. In Figure (\ref{fig:tmse}) the variation in the
entanglement of the system
and the entropy of the joint and subsystems is compared as the degree of
mixing is varied. This illustrates the result where the entanglement
of one of the pure states is not at a maximum and we see that the
entanglement of the joint state decreases as the
mixing is increased, with $\gamma = 0$ corresponding to the maximally
entangled pure state and $\gamma = 1 $ the non-maximally entangled
pure 
state. When the state is not maximally entangled the subsystems are not
totally mixed and hence the system tends towards the maximally entangled
pure state characteristics as the subsystem entropies tend towards $\log(2)$,
as $\gamma$ goes to zero. 
Figure (\ref{fig:tmsea}) illustrates where the
entropy of the subsystems are not always  equal and how the entanglement peaks
when the two subsystem entropies are both equal. The two turning
points where the entanglement goes to zero correspond to separable points, analogous to the case of equal
mixtures of maximally entangled components, where the entanglement
switches from a reliance on one  entangled component  to the other.

Already a great deal of complexity can be seen to be emerging from a
consideration of the entanglement and entropy characteristics of a
relatively simple system involving two polarisation-entangled qubits. This
is before the extended Hilbert space is introduced via the Beam Splitter
Protocol which itself introduces four new variables and hence a higher
degree of complexity again.

\begin{figure}[!h]
\begin{center}
\epsfig{figure=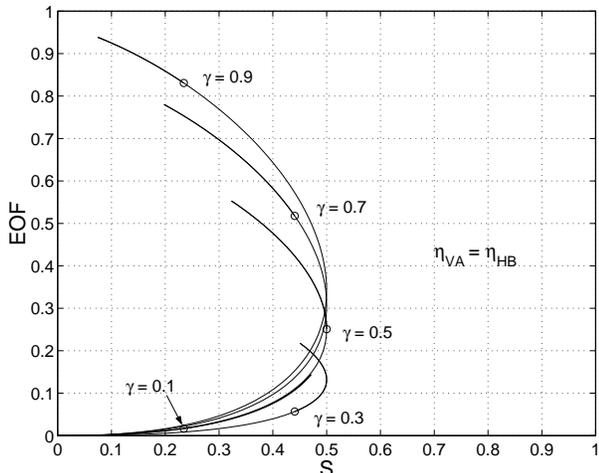,width=80mm}
\caption{\label{fig:tm23} This figure shows the EOF as a function of
the  entropy, S, of
the joint state,  for a range of states of the form of
eq.(\ref{eq:tmstate}), with variation dependent on the mixing and the
entanglement of the components. These involve a mixture of a
maximally entangled state and one that is weighted towards the $|VH\rangle$
modes. The circles indicate the characteristics for the
initial state and the solid line indicates the range of state
characteristics obtainable by varying two beam splitters in
unison. As the  bean splitter
transmission is reduced, the curve traces out  the
Concentration characteristics of the state until the maximum entanglement is
reached and the characteristics retrace their path and approach the
zero point.}
\end{center}
\end{figure}

For pure states the question of optimality of an entanglement
transformation for single-copy bipartite states is known. If
optimality of mixed states is taken as the most entanglement
that can be realised from a state regardless of the joint state entropy,
then the
conditions to obtain this can be found by looking at the local density
matrices of $A$ and $B$. The local density matrices are found by
tracing over the degrees of freedom for the other subsystem,
$\rho_A = {\rm Tr}_B\left[\rho_{AB}\right]$ and $\rho_B =
{\rm Tr}_A\left[\rho_{AB}\right]$. In
terms of this, the condition for a Bell diagonal, optimally entangled,
state is that $\rho_A = \rho_B = I_2$, corresponding to totally mixed, or
random, local
density matrices with a maximum amount of entropy, a condition
proposed by Kent {\it et al.} \cite{Kent:99}.
How do the entanglement-entropy
characteristics for the state vary under the influence of the Beam Splitter
Protocol? In Figure (\ref{fig:tm23}) the characteristics of a range of
states that are obtainable,
using the Beam Splitter Protocol, for a
given initial state are plotted. A state of the form of
eq.(\ref{eq:tmstate}) is employed as this can be varied with respect to the
mixture and the degree of entanglement of the two pure-state components,
which have already been considered.
The data points marked with a circle indicate the entanglement-entropy
characteristics of the initial state and the solid lines represent a range
of states that can be accessed by varying the beam splitters.
The two figures are for a range of initial states
determined by the mixing parameter, $\gamma$, which is labeled on each
curve. The degree of entanglement of one component is reduced
below that of a maximally entangled pure state. 
Consider the case of $\gamma = 0.7$ in Figure (\ref{fig:tm23}). This state
has a mixing which is weighted slightly towards the maximally entangled
component, and  the other component has only a small degree of entanglement. It
is the modes with higher probability of being realised in the less
entangled component that are targeted by the Beam Splitter Protocol,
$\eta_{VA}$ and $\eta_{HB}$, and hence the mixed state tends towards the
maximally entangled pure state component. The maximally entangled component
is never fully extracted in this instance as a result of the problem
inherent in most mixed state manipulation protocols in  that the
transformations affect all components of the mixture, not just the
component that needs to be removed. 

This protocol relies on a certain amount of {\it prior knowledge} of the
state. This knowledge helps determine the required parameters
to concentrate the state characteristics.  We consider concentration as
increasing both the entanglement and purity of a state\cite{Thew:01}.  Recent
advances in quantum state Tomography \cite{White:99} allow for the
measurement and reconstruction of the complete density matrix for a
bipartite state, allowing before and after comparisons.

 So far the manipulation
protocol has been introduced and shown to work for pure states, and it
has also been shown to increase both the entanglement and purity of a
state consisting of a mixture of pure states, one non-maximally entangled
and another
maximally entangled. The pure state transformations have been shown to
be optimal, both in the sense that the most amount of entanglement is
obtained by the transformation and the transformation is carried out
with an optimal efficiency. Can this be extended to arbitrary mixed states?
How this protocol works will constitute the majority of the next section, where this state, as well
as a range of other states, will be considered and an attempt made to
extrapolate some of the results to general systems. Chiefly, the beam
splitter dependence on the joint state and the subsystems will be
examined in greater detail to determine whether a state can be transformed,
and if it can, what beam splitters need to be varied, and by how much, to
obtain the most amount of entanglement for the state.


\section{Analysing Mixed States}

The main focus of this paper is the Polarising
Beam Splitter Protocol and mixed state entanglement, however, one of the reasons for
actually wanting such a device lies in its ability to explore mixed-state Hilbert space. To further illustrate the capabilities of the
protocol and at the same time investigate some recent proposals
concerning
entanglement and various concepts and bounds, a few states will be discussed in
detail. To describe the manipulations of a state and look at
questions regarding optimality and how these manipulations can be realised, it
will be necessary to look at the eigenvalues for the joint system
and local subsystems. In this
section a more detailed investigation into the mixed state already
introduced will firstly take place. The second state that will be
looked at will be the Werner state and then finally a state that is a
mixture of a pure entangled state and a separable component will be
introduced and discussed with a view to determining a bound on the
entanglement-entropy plane. A parameterised density matrix will
finally evolve from this, and then the Beam Splitter Protocol and the state
manipulations will be discussed again in the context of this bound.
\subsection{Two Bell State Mixture}
Consider again the state consisting of a mixture of two Bell-like states
\begin{eqnarray}
\hat{\rho}(\gamma) & = & \gamma |\phi^+\rangle\langle \phi^+| + (1-\gamma)|\psi^+\rangle\langle\psi^+|
\end{eqnarray}
which has the explicit density matrix after the beam splitter interaction of
\begin{eqnarray}
\tilde{\rho}  = & \frac{1}{P_B} & \left[\gamma \left( \begin{array}{cccc}
\eta_{VA}^2\eta_{VB}^2\cos^2\theta_1 & 0 & 0
&\bar{\eta} \cos\theta_1 \sin\theta_1  \\
0 & 0 & 0  & 0 \\
0 & 0 & 0  & 0 \\
\bar{\eta} \cos\theta_1 \sin\theta_1 & 0 & 0
&\eta_{HA}^2\eta_{HB}^2\sin^2\theta_1 \end{array} \right.\right)\nonumber \\
 &+ &(1-\gamma)\left.\left( \begin{array}{cccc}
0 & 0 & 0 & 0 \\
0 & \eta_{VA}^2\eta_{HB}^2\cos^2\theta_2 &
\bar{\eta}\cos\theta_2 \sin\theta_2 & 0 \\
0 & \bar{\eta} \cos\theta_2 \sin\theta_2 &
\eta_{HA}^2\eta_{VB}^2\sin^2\theta_2 & 0 \\
0 & 0 & 0 & 0 \end{array} \right) \right]
\end{eqnarray}
with $\bar{\eta} = \eta_{VA}\eta_{HA}\eta_{VB}\eta_{HB}$ as previously, and
where
\begin{eqnarray}
P_B &= &\gamma\left[\eta_{VA}^2\eta_{VB}^2\cos^2\theta_1 +
\eta_{HA}^2\eta_{HB}^2\sin^2\theta_1\right] \nonumber \\ & &+
~(1-\gamma)\left[\eta_{VA}^2\eta_{HB}^2\cos^2\theta_2 +
\eta_{HA}^2\eta_{VB}^2\sin^2\theta_2\right]
\end{eqnarray}
To determine the conditions to optimise the entanglement, the reduced
density operators for each of the subsystems need to be found. If the subsystem entropies are to be maximised, $S_A = S_B
= \log(2) $, then the following constraints must be satisfied
\begin{eqnarray}\label{eq:bbca}
\tan\theta_1 = \frac{\eta_{VA}\eta_{VB}}{\eta_{HA}\eta_{HB}}
\end{eqnarray}
\begin{eqnarray}\label{eq:bbcb}
\tan\theta_2 = \frac{\eta_{VA}\eta_{HB}}{\eta_{HA}\eta_{VB}}
\end{eqnarray}
As both the Entanglement of Formation and the Entropy are dependent on
the eigenvalues of the system, it is beneficial to determine how these
behave with regard to the
subsystem constraints. 

In satisfying the constraints on the local
systems, the eigenvalues for the joint system simplify to
\begin{eqnarray}
\lambda_1 & = & 2\gamma \eta_{VA}^2\eta_{VB}^2\cos^2\theta_1\frac{1}{P_B}\\
\lambda_2 &=& 2(1-\gamma)\eta_{VA}^2\eta_{HB}^2\cos^2\theta_2\frac{1}{P_B}
\end{eqnarray}
Given this, consider the ratio of these eigenvalues:
\begin{eqnarray}
\frac{\lambda_1}{\lambda_2} =
\left(\frac{\gamma}{1-\gamma}\right)\frac{\sin 2\theta_1}{\sin 2\theta_2}
\end{eqnarray}
Note that this ratio is independent of any transmission
coefficients only when the subsystem constraints are satisfied.
Thus by satisfying the subsystem constraints, the joint system requirements
are also being realised in that the degree of mixing of the state is
reduced as much as  possible, given the parameters governing the
initial state.  When this ratio equals one, the joint system is maximally mixed and
there is no entanglement present. Recall Figure (\ref{fig:tmse}), the entanglement  minima corresponds to the point where
$\lambda_1 / \lambda_2 = 1$. For a maximally entangled state this is at
$\gamma = 1/2$ but if $\theta \ne \pi/4$, then the minima will be
appropriately shifted.  Which one of the joint state eigenvalues will
dominate is determined by both the mixing and the entanglement of the
pure state components in the original mixture. 

These constraints govern how much the state can be improved by the beam
splitters.  In general, the entanglement of a state is reduced as the degree of
mixing is increased, and this provides a bound on the possible
transformations for the state. By satisfying the local system constraints
proposed by Kent {\it et al.}\cite{Kent:99}, regarding optimal entanglement
enhancement, the joint state eigenvalues obtain their optimal
value. 

\subsection{Werner State}
The Werner state can be considered as  a weighted mixture of all four of the
Bell states \cite{Werner:89}, a straightforward extension of the
mixture of two Bell states just discussed. However, we consider the
Werner state, in the form
\begin{eqnarray}
\hat{\rho}_w(\gamma) &= & (1-\gamma)~\frac{I_4}{4} + \gamma~|\phi^+\rangle\langle\phi^+|
\end{eqnarray}
where  the initial state, $|\phi^+\rangle$, has a probability $\gamma $ of being
transmitted without errors and there is a component $(1 - \gamma)$ of
a totally mixed state. In the case where $\gamma \le 1/3$ the
state is separable and as such has no entanglement to recover or maintain.
If the pure state component of this mixture is maximally entangled then
$S_A = S_B = \log(2)$ regardless of the degree  of mixing and the state is
as
entangled as it can be. It is, perhaps, this characteristic that suggested  it might
provide a bound on how entangled a mixed state could be. If, however, 
the pure state component is not maximally entangled, this constraint does
not hold.

If the Beam Splitter Protocol is now implemented on a Werner state,
with a non-maximally entangled pure state component, then the
constraints on the subsystem  and joint state eigenvalues  are
determined, as in the case of the mixture of two Bell states previously.
Firstly  the subsystems are considered and the constraints that maximise
the subsystem entropies are determined.  The requirement for the local
density matrices of the Werner state, post-beam splitters, to satisfy $S_A = S_B = \log(2)$  are 
\begin{eqnarray}
\eta_{VA} = \eta_{VB}, && \;\;\;\;\;\;\;\; \eta_{HA} = \eta_{HB} \label{eq:wc1}\\
 {\rm and}\;\;\;\;\;\;\;\; & & \;\;\;\;\;\;\;\;  \tan \theta  =  \frac{\eta^2_{VA}}{\eta^2_{HA}}\label{eq:wc2}
\end{eqnarray}
The joint system for the Werner state has four eigenvalues, which
can be simplified using the previous constraints, and the ratios of the eigenvalues are again independent of the beam splitter
coefficients, when the subsystem constraints are satisfied.

So, if the subsystem entropies can be made to equal $\log(2)$, the degree
of mixing in the joint state is minimised with respect to the maximum
amount of entanglement. Also, as there is no means of removing any of the eigenvalues, a relatively
high degree of mixing will be inherent in the system even after the Beam
Splitter Protocol. This mixing will however be the minimum obtainable
while maintaining $S_A = S_B = \log(2)$.
Immediately it can be seen, given the constraints of eq.(\ref{eq:wc1}) and
eq.(\ref{eq:wc2}), that if $\tan\theta \le 1$, then  by setting
$\eta^2_{HA} = \eta^2_{HB} = 1$ and  $\eta^2_{VA} = \eta^2_{VB} =
\tan\theta$,  the mixing and entanglement are optimised.

\begin{figure}[!htb]
\begin{center}
\epsfig{figure=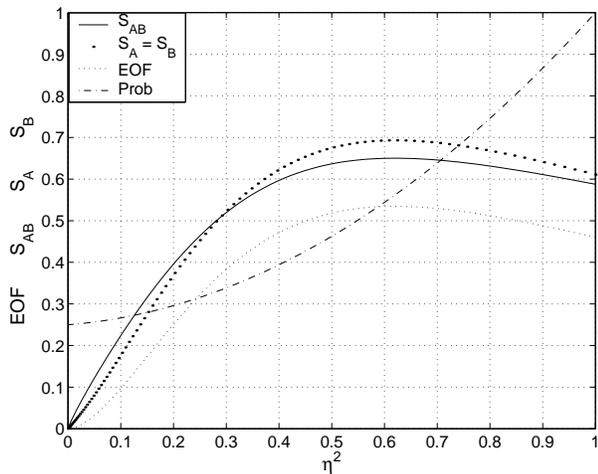,width=80mm}
\caption{\label{fig:wernera} The variation of the
entanglement and entropies, of joint and subsystems, for the Werner state, and the
Probability of obtaining these characteristics as two beam splitters are
varied in unison, $\eta^2 \equiv \eta_{VA}\eta_{VB}=\eta_{VA}^2$.}
\end{center}
\end{figure}

\begin{figure}[!h]
\begin{center}
\epsfig{figure=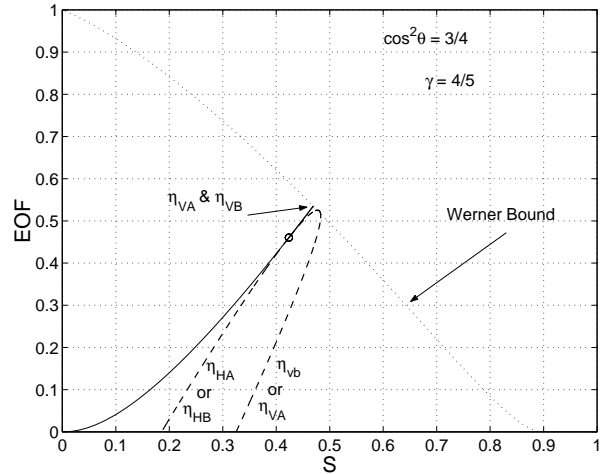,width=80mm}
\caption{\label{fig:wernerb} The $EOF$ and the entropy of the joint
system are shown for a range of Werner states, when the beam splitters are
varied individually and in unison. The peak value on the bounding
curve corresponds to the tuning parameters from Figure (\ref{fig:wernerb}) where $S_A = S_B = \log(2)$.}
\end{center}
\end{figure}

In Figure (\ref{fig:wernera}) the entropies and the entanglement of
the state are plotted as a function of the two transmission
coefficients, $\eta_{VA}$ and $\eta_{VB}$. These two beam splitters are varied
in unison, $\eta^2 \equiv \eta_{VA}\eta_{VB} =\eta_{VA}^2$,  to satisfy the
constraints on the subsystems. Given this, the
entanglement of the state is
maximised at the point where the entropy of the local subsystems
reaches a maximum, $\log(2)$. The probability of the state with these characteristics being
realised is also shown. When
the state is as entangled as it can be (here an increase of around
20\% is shown) it can be seen that the probability of obtaining this
state is around 55\%. 
The variation in the entanglement-entropy
characteristics, Figure (\ref{fig:wernerb}), shows the results of
individually varying the beam splitters, and the improvement that is
achieved when they are varied in tandem. $\eta_{VA}$ and $\eta_{VB}$, when
varied individually, increase the entropy at the cost of entanglement, but
when varied in unison the increase in the entanglement is greater. The maximum entanglement in Figure (\ref{fig:wernerb}) again
corresponds to the point where $\eta_{VA}^2=\eta_{VB}^2 = \tan\theta
\approx 0.6$. 

 A boundary curve has been introduced in  Figure (\ref{fig:wernerb}) (dotted line) onto the
characteristic plane denoting the bound alluded to earlier for the Werner
state, where the pure state component is maximally entangled. The
curve denotes the characteristics as the $\gamma$ is varied.  Regardless of whether the manipulations are made,
individually or
in unison, and how many beam splitters are utilised, the entanglement does not
exceed this bound.
  
\subsection{Entangled + Separable}
In the preceding sections, discussion has revolved around  mixtures of pure
non-maximally entangled Bell states. The first, a mixture of two, and the
second, a mixture of all four, and the behaviour of the characteristics of the
state under the Beam Splitter Protocol have been observed. To
observe the behaviour of a different class of state, a mixture of an entangled
pure state and a separable state will now be considered, with the
mixture having  the form
\begin{eqnarray}\label{eq:mems1}
\hat{\rho}_{es}(\gamma) & =& \gamma ~|\psi^+\rangle \langle \psi^+| + (1 - \gamma)|VV\rangle
\langle VV|
\end{eqnarray}
 There are only two eigenvalues for the joint system, $\lambda_1 =
\gamma, \;\lambda_2 = (1- \gamma)$, independent of $\theta$, however,
the eigenvalues for the subsystems are  dependent on both the mixing
and the entanglement of the entangled component.

 After the beam splitters, the eigenvalues for the subsystem and the
constraints on the  joint system are determined, resulting in
two requirements, the first being that
\begin{eqnarray}\label{eq:memsc1}
\tan^2\theta = \frac{\eta_{VA}^2\eta_{HB}^2}{\eta_{HA}^2\eta_{VB}^2}
\end{eqnarray}
which is very similar to those constraints found for previous states. The second constraint poses an
interesting problem, or perhaps it should be considered a
feature.  The second constraint requires that
\begin{eqnarray}\label{eq:memsc2}
\eta_{VA}^2\eta_{VB}^2 \rightarrow 0
\end{eqnarray}
This implies that the subsystems are totally mixed only in the limit
of no transmission.  Previously, when the subsystem constraints were enforced on the joint
system eigenvalues, the degree of mixing for the joint state had been
minimised. In this instance the ratio of the eigenvalues  is still
dependent on the beam splitter transmission coefficients. As such,
there is considerable control over the mixing of the state. This state
falls in to the class of state that was recently shown by Verstraete
{\it et al.} \cite{Verstraete:00} that could be brought arbitrarily close
to a Bell state by reducing the rank of the density operator.

\begin{figure}[!h]
\begin{center}
\epsfig{figure=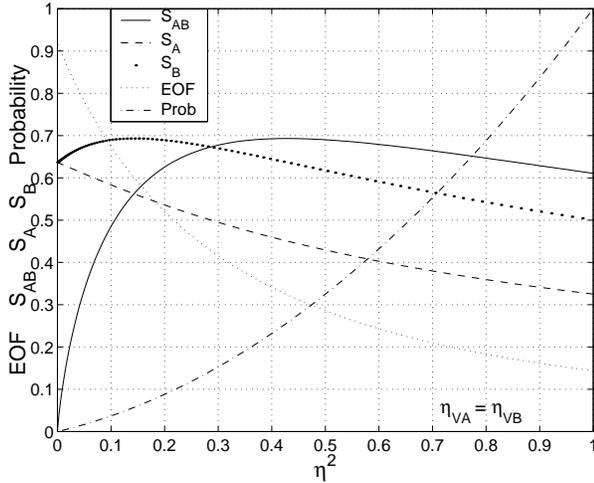,width=80mm}
\caption{\label{fig:eps12}The entanglement and entropy characteristics
for a mixture of non-maximally entangled and separable components as
the two beam splitters coefficients are varied with both equal ($\eta^2 \equiv \eta_{VA}\eta_{VB} = \eta^2_{VA}$). The subsystem entropies
are not equal except in the limit where the joint state
entropy initially increases and  only
decreasing as the transmission coefficients tend to zero. The pure
non-maximally entangled state characteristics are obtained in the limit as the
transmission goes to zero.}
\end{center}
\end{figure}

The behaviour and the entanglement-entropy
characteristics of this state, as the beam splitters are varied, is
illustrated in Figure (\ref{fig:eps12}). The figure highlights the
effects that both of these constraints have on the state
characteristics.
The mixture consists of $ \gamma = 0.3$ of the
entangled component and shows the results in the case where the two beam
splitters are varied together, $\eta_{VA} = \eta_{VB}$. With these
parameter settings the entropies of the
subsystems are not equal, $S_A \ne S_B$,
and only in the limit as the transmission coefficients both go to zero
do they converge, satisfying eq.(\ref{eq:memsc2}), and when they do, they are not a maximum and hence
the state is not maximally entangled.

If the first constraint, eq.(\ref{eq:memsc1}), is satisfied and
the transmission coefficients of the beam splitters are varied as $\eta_{VA} =
\eta_{VB}\tan\theta, \;(\tan\theta \le 1)$ then the behaviour is not
that different from this case, except that the subsystem entropies are equal
throughout the variation of the beam splitters and in the limit as the
transmission tends to zero, the maximally entangled pure state
characteristics are approached as $S_A = S_B$ approaches $\log(2)$. As this is achieved, the probability of obtaining these state
characteristics tends to zero.

\begin{figure}[!htb]
\begin{center}
\epsfig{figure=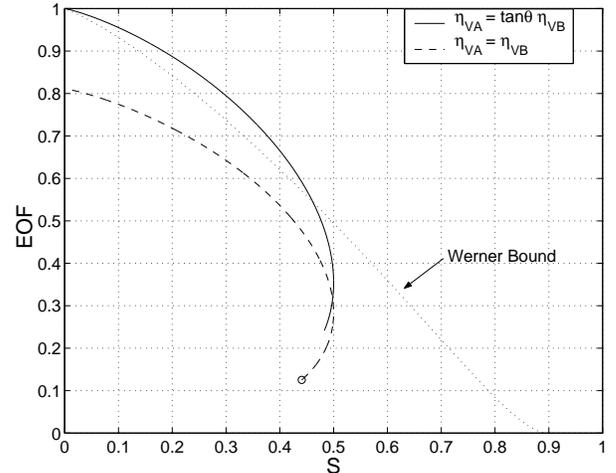,width=80mm}
\caption{\label{fig:eps34}The range of state characteristics
obtainable by implementing the Beam Splitter Protocol on a mixture of
entangled and separable states. The two curves, dashed and solid
respectively denote varying the beam splitters equally, and when the
subsystem constraints are satisfied. The latter case shows that the
characteristics can approach those of  a maximally
entangled state. The
Werner state bound is also shown and is crossed for some of the
states obtained. }
\end{center}
\end{figure}

In Figure (\ref{fig:eps34}) the state is again considered and the
behaviour of the characteristics on the entanglement-entropy plane
examined. The dashed curve denotes the behaviour as the beam splitters
are varied in tandem as in Figure (\ref{fig:eps12}), showing the pure
but non-maximally entangled state characteristics being approached. The
solid line shows the transformed states obtained by varying the beam
splitters while satisfying both the subsystem constraints. This curve
shown doesn't  pass through the circle marking the initial state
characteristics, due to the fact that the initial state, numerically
considered, has some $\eta$ dependence such that initial $\eta \ne 1$, again due to the first constraint.
This state has already presented some quite unusual characteristics, however
it can also be noted that for variations in the mixture  and entanglement
of this state, the same concentration characteristics as in Figure
(\ref{fig:eps34}) are produced. In fact, this solid curve can be further
extrapolated
down the plane,  for as long as there is some component of the entangled
pure state present in the mixture, a state arbitrarily close to a maximally entangled state can be
recovered. The probability of such a situation is proportionally unlikely.
In the case of a third of the mixture having entanglement corresponding to
a 1/3:2/3, weighting of the entangled pure state components, the
probability of a maximally entangled pure state being recovered is of the
order of
$10^6$. Regardless of this, this  state would
provide an incredibly useful state to construct with a view to experimentally
considering questions regarding mixed state quantum information,
in that it covers so much of the entanglement-entropy plane.

The characteristics of the Werner state suggested that it might
provide a bound on mixed state entanglement. In this figure, in the
case where the first subsystem constraint is satisfied, there exist
states with characteristics above  the Werner bound. Clearly the Werner state does not provide an absolute bound for mixed
state entanglement and a higher bound needs to be found.
\section{Bounds on Entanglement}
More recently, an attempt to put a bound on mixed state entanglement
has resulted in a proposal by Munro {\it et al.} \cite{Munro:00}. The
proposed bound involves a density matrix, not entirely dissimilar to
the previous mixture of an entangled and a separable state. By
considering a slight variation on this state, which involves placing
some restrictions on the density matrix elements, a state of the form
\begin{eqnarray}\label{eqn:Hin}
\hat{\rho}(\gamma)& = & \left( \begin{array}{cccc}
            1 - 2g(\gamma) & 0 & 0 & 0 \\
	            0 & g(\gamma) & \gamma/2 & 0 \\
		0 &\gamma/2  & g(\gamma) & 0 \\
	     0  & 0 & 0 & 0  \\ \end{array} \right)
\end{eqnarray}
is obtained, where $g(\gamma) = \gamma/2$ for $\gamma \ge 2/3$ and
$g(\gamma) = 1/3$ for $ \gamma < 2/3$. 

This state has a maximum amount
of entanglement for the degree of purity, in terms of the Linear
Entropy where the Linear Entropy \cite{Bose:99a}, $S_L = 1 - P$, is
related to the purity of the state, $P = {\rm Tr} [\rho^2]$. This is
then normalised  so that
\begin{eqnarray}
S_L &=& \frac{4}{3} \left( 1 - {\rm Tr} [\rho^2] \right)
\end{eqnarray}
returns a value ranging from 0 for pure states, to 1 for a totally
mixed state. This state has very similar behaviour to the previous
state for $\gamma \ge 2/3$ but significantly different below this
point. For this state there are two non-zero eigenvalues for $\gamma > 2/3$ and
three below this point. Note here that $\gamma$ is not the mixture coefficient as
previously defined. 

If the subsystem constraints are again considered, the following
restriction apply
\begin{eqnarray}\label{eq:slbc1}
\eta_{VA}^2 \eta_{HB}^2 & = &  \eta_{HA}^2 \eta_{VB}^2
\end{eqnarray}
and as before
\begin{eqnarray}
\eta_{VA}^2 \eta_{VB}^2 \rightarrow 0
\end{eqnarray}
 It is the
case where  $\gamma < 2/3$ that the state has the most significant
change in its
entanglement-entropy characteristics.  Below $\gamma =2/3$ the
behaviour of this bounding state differs markedly from that of the
previous state. The emergence of the extra eigenvalue increases the
entropy and hence extends the coverage of the bound on the
entanglement-entropy plane.  As such this would suggest
that there is some higher bound with respect to the Entanglement of
Formation and entropy above what this state proposes. This might also suggest that if the bound is
going to be complete, then at some point the emergence of a bounding
state with four eigenvalues may be necessary as the entanglement tends
to zero. At this point the Werner state may indeed provide the
small entanglement - large mixing bound. 
\begin{figure}[!htb]
\begin{center}
\epsfig{figure=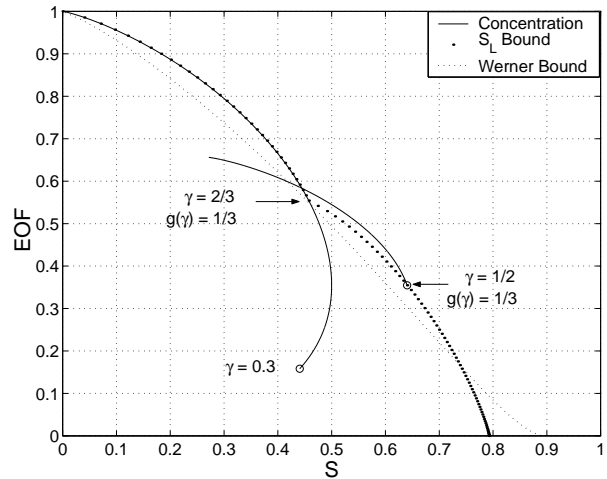,width=80mm}
\caption{\label{fig:mixprob1}The $EOF$ and entropy characteristics,
solid line, for the state given by eq.(\ref{eq:mems1}) starting at
$\gamma = 0.3$ and the Linear Entropy bound state at $g(\gamma) =
1/2$.  The first state has the feature that it approaches and then
follows the linear entropy bounding curve up to the peak value. It
coincides with the Linear Entropy curve at $\gamma =2/3$. Both the Werner
and linear
entropy bounds are shown and the Beam Splitter protocol enables the
state characteristics to
exceed the Linear Entropy bound. An example of a state starting at $\gamma
= 1/2$ is shown.}
\end{center}
\end{figure}
In Figure (\ref{fig:mixprob1}) the Werner state bound
and the Linear Entropy bound are shown, and as just suggested, as
the entanglement approaches zero, the Werner state bound is greater, going to
zero at $\gamma = 1/3$, $S \sim 0.9$, as the state becomes separable.  The
previous state  is shown again here, where
a mixture containing a maximally entangled pure state is used. This
state and the Werner State both  coincide with the 
Linear Entropy bound  at $\gamma
= 2/3$.  The previous state covers quite a large region of this space
with the  characteristics first increasing in entanglement at the cost of
entropy before both the entanglement and the entropy
improve, and concentration is realised. 

The other curve on this figure to notice is the solid line starting at
$\gamma = 1/2$. This curve denotes the characteristics of the Linear
Entropy bounding state when the Beam Splitter Protocol is applied to
it. Notice that the boundary curve is exceeded, both above and
below the bifurcation point for the state at $\gamma = 2/3$. This
should not be unexpected, as the subsystem entropies are not
maximised for this state. The state was optimised in terms of the
Linear Entropy of the joint state. Optimisation in terms of the von
Neumann entropy is currently underway. 

\section{Discussion}

There have been several key points looked at in this paper and all of
these have revolved around the Beam Splitter
Protocol for manipulating mixed states. In doing this the equivalence
between the coincidence detection state and that obtained by a ``detect
and discard'' protocol, with perfect detectors at the reflected ports
of the beam splitters, has also been justified. The process was then
re-introduced, equivalently, in the context of local filtering operations. Although
only a limited range of states were considered, the way the  Beam Splitter Protocol transforms
a state has shown that, due to the large number of degrees of freedom of the
protocol, the scheme is highly adaptable. The transformations have been shown
to extract the most amount of entanglement from a state  that is
possible for a given degree
of mixing and in this sense could be considered optimal.
The question of a bound on the amount of entanglement that a mixed state
can have has been explored, firstly the Werner state and then the Linear
Entropy bound. 

For mixed states, questions of efficiency and optimality are not
 clear and as such discussion regarding these have been limited. The distinction is  made regarding these concepts -
 it is one thing for the transformation to obtain the final state with
 the most amount of entanglement that is possible, and it is another  to show
the optimal probability or efficiency of carrying out a particular state
transformation.
The proposed bound on entanglement enhancement of Kent {\it et al.}
\cite{Kent:99} applies
to the first interpretation and the protocol is shown to satisfy these
 requirements.  In the case of the second interpretation, there has recently been a
proposal by Vidal \cite{Vidal:00} which is an extension of his ideas on
single-copy
pure states and requires a minimisation over a set of
entanglement monotones which has been left for future work.

The primary piece of information to note here is that:  IF there is some initial
amount of entanglement, AND the subsystems are not BOTH totally mixed, THEN
more entanglement can be obtained by transforming the mixed state. The Beam
Splitter Protocol  introduced here  can achieve this, it can do
it in a very simple way, and one that is experimentally realisable.

 The recovery and maintenance of an
entanglement resource is a process that will be of  paramount importance
for any form of reliable quantum communication. Just as
important is the
investigation of mixed state entanglement in its own right. An
understanding of the
relationship between the classical and quantum probability
distributions specifying an entangled mixed state is still not complete and
any opportunity to  investigate this  in an
experimental regime should be promoted.

The authors would like to thank A.G. White, P.G. Kwiat, K. Nemoto,
M. Nielsen and G.J. Milburn for helpful discussions and support.  We also
thank S. Massar for bringing to our attention ref\cite{Verstraete:00}.

\end{multicols}
\end{document}